\documentclass[runningheads]{llncs}
\usepackage{amssymb,amsmath}
\usepackage{pstricks}
\usepackage{pst-node}
\usepackage{epsfig}
\usepackage{latexsym}
\begin{document}
\pagestyle{headings}

\setlength{\unitlength}{1mm}

\newcommand{\xnot}   {{\tt not} \ }
\newcommand{\xIn}    {\mathsf{Insert}}
\newcommand{\xOut}   {\mathsf{Retract}}
\newcommand{\xCWA}   {\mathsf{CWA}}
\newcommand{\xDB}    {({\cal D},\,{\cal IC})}
\newcommand{\DB}     {\ensuremath{{\cal DB}}}
\newcommand{\UDB}    {\ensuremath{{\cal UDB}}}

\newcommand{\asystem} {${\cal A}$\mbox{-system }}

\titlerunning{Repairing Inconsistent Databases}
\title     {Repairing Inconsistent Databases: \\
            A Model-Theoretic Approach \\
            and Abductive Reasoning
\thanks{Originally published in proc. PCL 2002, a FLoC workshop;
eds. Hendrik Decker, Dina Goldin, J{\o}rgen Villadsen, Toshiharu Waragai
({\tt http://floc02.diku.dk/PCL/}).}
}
\author    {Ofer Arieli\inst{1} \and
            Marc Denecker\inst{2} \and
            Bert Van Nuffelen\inst{2} \and
            Maurice Bruynooghe\inst{2}}
\institute {Department of Computer Science,
            The Academic College of Tel-Aviv \\
            Antokolski 4, Tel-Aviv 61161, Israel \\
            \email{oarieli@mta.ac.il}
            \and
            Department of Computer Science,
            The Catholic University of Leuven \\
            Celestijnenlaan 200A, B-3001 Heverlee, Belgium \\
            \email{$\{$marcd$,$bertv$,$maurice$\}$@cs.kuleuven.ac.be}}
\maketitle

\begin{abstract}
In this paper we consider two points of views to the problem of
coherent integration of distributed data. First we give a pure
model-theoretic analysis of the possible ways to `repair' a
database. We do so by characterizing the possibilities to
`recover' consistent data from an inconsistent database in terms
of those models of the database that exhibit as minimal
inconsistent information as reasonably possible. Then we
introduce an abductive application to restore the consistency of
a given database. This application is based on an abductive
solver (${\cal A}$-system) that implements an SLDNFA-resolution
procedure, and computes a list of data-facts that should be
inserted to the database or retracted from it in order to keep
the database consistent. The two approaches for coherent data
integration are related by soundness and completeness results.
\end{abstract}

\section{Introduction}

Integration of data coming from different databases is a very
common, nevertheless nontrivial, task. There are a number of
different phases involved in this process, the most important of
which are the following:
\begin{enumerate}
   \item Resolving the different ontologies and/or database scheme,
         setting a single unified schema, and translating the
         integrity constraints\footnote {I.e., the rules that represent
         intentional truths of a database domain.} of each database to
         the new ontology.
   \item Resolving contradictions among the integrity constraints of
         different local databases.
   \item Integrating distributed databases w.r.t. the unified set of
         integrity constraints, computed in the previous phase.
\end{enumerate}

Each one of the phases mentioned above has its own difficulties
and challenges. For instance, we are not aware of any work that
gives a complete and robust solution to the problem of the first
phase. Most of the formalisms for database integration implicitly
assume that all the databases to be integrated have the same
ontology, so the first phase is not needed.

The reason for separating the remaining two phases is that
integrity constraints represent truths that should be valid in
all situations, while a database instance represents an
existentional truth, i.e., an actual situation. Consequently, the
policy of resolving contradictions among integrity constraints is
often different than the one that is applied on database facts,
and the former should be applied first.

Despite their different nature, both these phases are based on
some formalisms that maintain contradictions and allow to draw
plausible conclusions from inconsistent situations. Roughly,
there are two approaches to handle this problem:
\begin{itemize}
   \item {\em Paraconsistent\/} formalisms, in which the amalgamated data
         may remain inconsistent, but the set of conclusions implied by it
         is not explosive, i.e.: not every fact follows from an
         inconsistent database.
         Paraconsistent procedures for integrating data (e.g.,
         \cite{dACM02,Su94}) are often based on a paraconsistent reasoning
         process, such as LFI \cite{CM01}, annotated logics \cite{KL92,Su90},
         or other non-classical proof systems \cite{AA96,Pr91}.
   \item {\em Coherent\/} (consistency-base) methods, in which the
         amalgamated data is revised in order to restore consistency (see,
         e.g., \cite{AA99,BKM91,BS02,GZ00,LS00}). In many cases the
         underlying formalism of these approaches are closely related to the
         theory of belief revision \cite{AGM95,GH95}. In the context of
         database systems the idea is to construct consistent databases that
         are ``as close as possible'' to the original database. These
         ``repaired'' instances of the spoiled database correspond to
         plausible and compact ways of restoring consistency.
\end{itemize}

In this paper we follow the latter approach, and consider two
points of views for the last phase of the process, namely:
coherent methods of integrating distributed databases (with the
same ontology) w.r.t. a consistent set of integrity constraints.
The main difficulty in this process stems from the fact that even
when each local database is consistent, the collective
information of all the distributed databases may not be consistent
anymore. In particular, facts that are specified in a particular
database may violate some integrity constraints defined
elsewhere, and so it might contradict some elements in the
unified set of integrity constraints. Our goal is therefore to find
ways to properly ``repair'' a combined database, and restore its
consistency.

One way of viewing this problem is by a model-theoretic analysis
that characterizes database repairs in terms of a certain set of
models of the inconsistent database (those that, intuitively,
minimize the amount of inconsistent information). The other
approach is based on abductive reasoning. For this we use an
abductive solver (${\cal A}$-system, \cite{KvND01}) that
implements SLDNFA-resolution \cite{DD92,DD98} for computing a
list of data-facts that should be inserted to the database or
retracted from it in order to keep the data consistent. A
corresponding application was introduced and described in greater
details in \cite{AvNDB01}. Here we review it in order to keep this
paper self contained, and putting our results in the right
context. We then show that the abductive process of coherent
integration of databases is sound and complete w.r.t. the
semantics that is induced by the model theoretic analysis.
\footnote{Due to a lack of space some proofs are reduced or omitted
altogether. Full proofs will appear in an extended version of this paper.}

\section {Coherent integration of databases}

In this paper we assume that we have a first-order language $L$,
based on a fixed database schema $S$, and a fixed domain $D$.
Every element of $D$ has a unique name. A database instance
${\cal D}$ consists of atoms in the language $L$ that are
instances of the schema $S$. As such, every instance ${\cal D}$
has a finite active domain, which is a subset of $D$. A {\em
database\/} is a pair $\xDB$, where ${\cal D}$ is a database
instance, and ${\cal IC}$, the set of {\em integrity
constraints\/}, is a finite set of formulae in $L$ (assumed to be
satisfied by ${\cal D}$).

Given a database $\DB \!=\! \xDB$, we apply to it the closed word
assumption, so only  the facts that are explicitly mentioned in
${\cal D}$ are considered true. The underlying semantics
corresponds, therefore, to minimal Herbrand interpretations.

\begin{definition}
{\rm The {\em minimal Herbrand model\/} ${\cal H}^{\cal D}$ of a
database instance ${\cal D}$ is the model of ${\cal D}$ that
assigns true to all the ground instances of atomic formulae in
${\cal D}$, and false to all the other atoms.
} \end{definition}

\begin{definition}
{\rm A formula $\psi$ {\em follows\/} from a database instance
${\cal D}$ (notation: ${\cal D} \models \psi$) if the minimal
Herbrand model of ${\cal D}$ is also a model of $\psi$. }
\end{definition}

\begin{definition}
{\rm A database $\DB \!=\! \xDB$ is {\em consistent\/} if ${\cal
IC}$ is a classically consistent set, and each formula of it
follows from ${\cal D}$ (notation: ${\cal D} \models {\cal IC}$).
} \end{definition}

Our goal is to integrate $n$ consistent databases, $\DB_{i} \!=\!
({\cal D}_{i},\,{\cal IC}_{i})$, $i \!=\! 1,\ldots n$, in such a
way that the combined data will contain everything that can be
deduced from one source of information, without violating any
integrity constraint of another source. The idea is to consider
the union of the distributed data, and then to restore its
consistency. A key notion in this respect is the following:

\begin{definition}
\label{def:repair}
{\rm A {\em repair\/} of $\DB \!=\! \xDB$ is a
pair $(\xIn,\xOut)$ such that (1) $\xIn \cap {\cal D} \!=\!
\emptyset$,  (2) $\xOut \subseteq {\cal D}$,\footnote {Note that
by conditions (1) and (2) it follows that $\xIn \cap \xOut \!=\!
\emptyset$.} and (3) $({\cal D} \cup \xIn \setminus \xOut,\;{\cal
IC})$ is a consistent database.
} \end{definition}

Intuitively, $\xIn$ is a set of elements that should be inserted
into ${\cal D}$ and $\xOut$ is a set of elements that should be
removed from ${\cal D}$ in order to obtain a consistent database.

\begin{definition}
\label{def:repairedKB}
{\rm A {\em repaired database\/} of $\DB
\!=\! \xDB$ is a consistent database $({\cal D} \cup \xIn
\setminus \xOut \, , \, {\cal IC})$, where $(\xIn,\xOut)$ is a
repair of $\DB$.
} \end{definition}

As there may be many ways to repair an inconsistent
database,\footnote {Some of them may be trivial and/or useless.
For instance, the inconsistency in $\xDB = (\{p,q,r\},\{\neg p\})$
may be removed by deleting every element in ${\cal D}$, but this
is certainly not the optimal way of restoring consistency in this
case.} it is often convenient to make preferences among the
possible repairs, and consider only the most preferred ones.
Below are two common preference criteria.

\begin{definition}
\label{preference_criteria}
{\rm Let $(\xIn,\xOut)$ and $(\xIn',\xOut')$ be two repairs.
\begin{itemize}
   \item {\em set inclusion preference criterion \/}:
         $(\xIn',\xOut') \leq_{i} (\xIn,\xOut)$, if
         $\xIn \subseteq \xIn'$ and $\xOut \subseteq \xOut'$.
   \item {\em cardinality preference criterion\/}:
         $(\xIn',\xOut') \leq_{c} (\xIn,\xOut)$ if
         $|\xIn| + |\xOut| \!\leq\! |\xIn'| + |\xOut'|$.
         \footnote {Set inclusion is also considered in
         \cite{ABC99,BS02,dACM02,GZ00}; cardinality is considered,
         e.g., in \cite{LS00}}
\end{itemize}
} \end{definition}

In what follows we assume that $\leq$ is a fixed pre-order that represents
some preference criterion on the set of repairs.

\begin{definition}
\label{def:preferred_repairs}
{\rm A {\em $\leq$-preferred
repair\/} of  $\DB$ is a repair $(\xIn,\xOut)$ of $\DB$, s.t. for
every repair $(\xIn',\xOut')$ of $\DB$, if $(\xIn,\xOut) \!\leq\!
(\xIn',\xOut')$ then $(\xIn',\xOut') \!\leq\! (\xIn,\xOut)$. The
set of all the $\leq$-preferred repairs of $\DB$ is denoted by
$!(\DB,\leq)$. }
\end{definition}

\begin{definition}
{\rm A {\em $\leq$-repaired database\/} of $\DB$ is a repaired
database of $\DB$, constructed from a {\em $\leq$}-preferred
repair of $\DB$. The set of all the $\leq$-repaired databases of
$\DB$ is denoted by \vspace{1mm} \\
\hspace*{2mm} ${\cal R}(\DB,\leq) = \{ \, ({\cal D} \cup \xIn
\setminus \xOut \, , \, {\cal IC}) ~|~ (\xIn,\xOut) \in
\!(\DB,\leq) \, \}$. }
\end{definition}

Note that if $\DB$ is consistent, and the preference criterion is
a partial order that is monotonic in the total size of the
repairs' components (as in Def. \ref{preference_criteria}), then
${\cal R}(\DB,\leq) \!=\! \{\DB\}$, so
there is nothing to repair,
as expected.

It is usual to refer to the $\leq$-preferred databases of $\DB$
as the consistent databases that are `as close as possible' to
$\DB$ itself (see, e.g., \cite{ABC99,dACM02,LS00}). Indeed,
denote $Th({\cal D}) = \{P(t) ~|~ {\cal D} \models P(t) \}$,
where $P$ is a relation name and $t$ is a ground tuple, and let
${\tt dist}({\cal D}_{1},{\cal D}_{2})$ be the following set:
\vspace{1mm} \\
\hspace*{12mm} ${\tt dist}({\cal D}_{1},{\cal D}_{2}) =
         (Th({\cal D}_{1}) \setminus Th({\cal D}_{2})) \cup
         (Th({\cal D}_{2}) \setminus Th({\cal D}_{1}))$
\vspace{1mm} \\
It is easy to see that $\DB' = ({\cal D}',{\cal IC})$ is a
$\leq_{i}$-repaired database of $\DB = \xDB$, if the set ${\tt
dist}({\cal D}',{\cal D})$ is minimal (w.r.t. set inclusion)
among all the sets of the form ${\tt dist}({\cal D}'',{\cal D})$,
where ${\cal D}'' \models {\cal IC}$. Similarly, if $\#(S)$
denotes the number of elements in $S$, then $\DB' = ({\cal
D}',{\cal IC})$ is a $\leq_{c}$-repaired database of $\DB =
\xDB$, if $\#({\tt dist}({\cal D}',{\cal D})$ is minimal in $\{
\#({\tt dist}({\cal D}'',{\cal D})) ~|~ {\cal D}'' \models {\cal
IC} \}$.

\begin{definition}
{\rm For $\DB_{i} = ({\cal D}_{i},\,{\cal IC}_{i})$, $i =
1,\ldots n$, let $\UDB = \xDB$, where ${\cal D} =
\bigcup_{i=1}^{n}{\cal D}_{i}$ and ${\cal IC} =
\bigcup_{i=1}^{n}{\cal IC}_{i}$. } \end{definition}

Given $n$ distributed databases and a preference criterion
$\leq$, our goal is to compute the set ${\cal R}(\UDB,\leq)$ of
the $\leq$-repaired databases of $\UDB$ (or to be able to
compute, in an efficient way, some elements in this set). Below
are test-cases for such database integration. \footnote {See,
e.g., \cite{ABC99,BS02,GZ00} for more discussions on the examples
below.} \footnote{In all the following examples we use set
inclusion as the preference criterion. In what follows we shall
fix a preference criterion for choosing the ``best'' repairs and
omit its notation whenever possible.}

\begin{example}
\label{example:teachers}
{\rm Consider a distributed database with a relation $teaches$ of the
 following scheme: $({\tt course\_name}, {\tt teacher\_name})$.
 Suppose also that each database contains a single integrity
 constraint, stating that the same course cannot be taught by two
 different teachers:  \vspace{1mm} \\
 \hspace*{5mm}
 ${\cal IC} = \{ \,
   \forall X \forall Y \forall Z \,
   (teaches(X,Y) \wedge teaches(X,Z) \, \rightarrow \,
   Y = Z ) \, \}$.
 \vspace{1mm} \\
Consider now the following two databases:
 \vspace{2mm} \\
 \hspace*{5mm}
 $\DB_{1} = ( \, \{teaches(c_{1},n_{1}), \:
                   teaches(c_{2},n_{2})\}, \, {\cal IC} \,)$, \vspace{1mm} \\
 \hspace*{5mm}
 $\DB_{2} = ( \, \{teaches(c_{2},n_{3})\}, \, {\cal IC})$ \vspace{1mm} \\
 Clearly, the unified database $\DB_{1} \cup \DB_{2}$ is inconsistent.
 Its preferred repairs are $(\emptyset, \,\{teaches(c_{2},n_{2})\})$ and
 $(\emptyset, \,\{teaches(c_{2},n_{3})\})$.
 Hence, the two repaired databases are the following: \vspace{1mm} \\
 \hspace*{5mm}
 ${\cal R}_{1} = ( \, \{teaches(c_{1},n_{1}), \: teaches(c_{2},n_{2})\},
                   \, {\cal IC} \, )$, \vspace{1mm} \\
 \hspace*{5mm}
 ${\cal R}_{2} = ( \, \{teaches(c_{1},n_{1}), \: teaches(c_{2},n_{3})\},
                   \, {\cal IC} \, )$.
} \end{example}

\begin{example}
\label{example:set_inclusion}
{\rm Let ${\cal D}_{1} \!=\! \{p(a),
 \: p(b)\}, {\cal D}_{2} \!=\! \{q(a), \: q(c)\}$, and ${\cal IC}
 \!=\! \{ \forall X (p(X) \!\rightarrow\! q(X)) \}$. Again,
 $({\cal D}_{1}, \emptyset) \cup ({\cal D}_{2}, {\cal IC})$ is
 inconsistent. The corresponding preferred repairs are $(\{q(b)\},
 \,\emptyset)$ and $(\emptyset, \,\{p(b)\})$.
 The repaired databases are
 ${\cal R}_{1} = 
\\ 
 ( \, \{p(a), \: p(b), \: q(a), \: q(b),
                   \: q(c) \}, \, {\cal IC} \, )$ and
 ${\cal R}_{2} = ( \, \{p(a), \: q(a), \: q(c) \}, \, {\cal IC} \, )$.
} \end{example}

\section {Database repair -- A model-theoretic point of view}

In this section we characterize the repairs of a given database in
terms of its models. First, we consider arbitrary repairs, and
show that they can be represented
either by {\em two-valued models\/} of the theory of integrity
constraints, or by {\em three-valued models\/} of the set of
integrity constraints and the set of literals, obtained by
applying the closed world assumption on the database facts. Then
we focus on the most preferred repairs, and show that a certain
subset of the three-valued models considered above can be used for
characterizing $\leq$-preferred repairs.

\begin{definition}
{\rm Given a valuation $\nu$ and a truth value $x$. Denote:
\vspace{1mm} \\ \hspace*{16mm}
 $ \nu^{x} = \{ p ~|~ p \mbox { is an atomic formula, and } \nu(p) = x \}$.
   \footnote {Note, in particular, that $({\cal H}^{\cal D})^{t} = {\cal
   D}$.}
} \end{definition}

The following two propositions characterize repairs in terms of two-valued
structures.

\begin{proposition}
\label{prop:repair_0a}
 Let $\xDB$ be a database and let $M$ be a two-valued model of
 ${\cal IC}$. Let $\xIn = M^{t} \setminus {\cal D}$ and $\xOut =
 {\cal D} \setminus M^{t}$. Then $(\xIn,\xOut)$ is a repair of
 $\xDB$.
\end{proposition}

\noindent{\em Proof:} The definitions of $\xIn$ and $\xOut$
immediately imply that $\xIn \cap {\cal D} \!=\! \emptyset$ and
$\xOut \!\subseteq\! {\cal D}$. For the the last condition in
Definition \ref{def:repair}, note that in our case ${\cal D} \cup
\xIn \setminus \xOut = {\cal D} \cup (M^{t} \setminus {\cal D})
\setminus ({\cal D} \setminus M^{t}) = M^{t}$. It follows that
$M$ is the least Herbrand model of ${\cal D} \cup \xIn \setminus
\xOut$ and it is also a model of ${\cal IC}$, therefore ${\cal D}
\cup \xIn \setminus \xOut \models {\cal IC}$. \hfill$\Box$

\begin{proposition}
\label{prop:repair_0b}
 Let $(\xIn,\xOut)$ be a repair of a database $\xDB$. Then there is a
 classical model $M$ of ${\cal IC}$,\footnote{Recall that we assume that
 ${\cal IC}$ is classically consistent, thus it has classical models.} such
 that $\xIn = M^{t} \setminus {\cal D}$ and $\xOut = {\cal D} \setminus M^{t}$.
\end{proposition}

\noindent{\em Proof:} Consider a valuation $M$, defined for every atom $p$
as follows: \vspace{2mm} \\
\hspace*{24mm} $M(p) = \left\{ \begin{array}{ll}
      t  \ \ & \mbox{ if $p \!\in\! {\cal D} \cup \xIn \setminus \xOut$, }
               \vspace{1mm} \\
      f  \ \ & \mbox{ otherwise. } \\
\end{array} \right.$ \vspace{2mm} \\
By its definition, $M$ is a minimal Herbrand model of ${\cal D}
\cup \xIn \setminus \xOut$. Now, since $(\xIn,\xOut)$ is a repair
of $\xDB$, we have that ${\cal D} \cup \xIn \setminus \xOut
\!\models\! {\cal IC}$, thus $M$ is a (two-valued) model of
${\cal IC}$. Moreover,
$\xIn \cap {\cal D} \!=\! \emptyset$ and $\xOut
\!\subseteq\! {\cal D}$, hence we have the following: \vspace{1mm} \\
$\bullet$ $M^{t} \setminus {\cal D} =
          ({\cal D} \cup \xIn \setminus \xOut) \setminus {\cal D} = \xIn$,
          \vspace{1mm} \\
$\bullet$ ${\cal D} \setminus M^{t} =
          {\cal D} \setminus ({\cal D} \cup \xIn \setminus \xOut) =
          \xOut$. \hfill$\Box$

\bigskip

The above formalization in terms of two-valued models has the
drawback that a unified database $\UDB$ in need of a repair is
inconsistent. In order to avoid reasoning on inconsistent
theories, and since classical logic can infer everything from an
inconsistent theory, we develop another formalization, based on a
three-valued semantics. The benefit of this is that, as we show
below, {\em any\/} database has models w.r.t. appropriate
three-valued semantics, from which it is possible to pinpoint the
inconsistent information, and thus it is also possible to extract
repairs for $\UDB$.

\begin{figure}[hbt]
\begin{picture}(50,37)(-36,0)
\put(0,0){\vector(0,1){37}}    \put(01,35){$\leq_{k}$}
\put(0,0){\vector(1,0){50}}    \put(48,02){$\leq_{t}$}
\put(05,10){\circle*{2}}       \put(04,05){$f$}
\put(25,30){\circle*{2}}       \put(24,33){$\top$}
\put(45,10){\circle*{2}}       \put(44,05){$t$}
\put(05,10){\line(1,1){20}}
\put(25,30){\line(1,-1){20}}
\end{picture}
\caption{The structure ${\cal THREE}$}
\label{fig:three}
\end{figure}
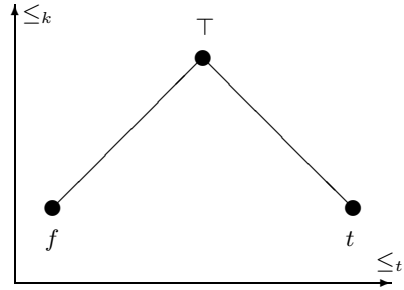

The underlying 3-valued semantics considered here is induced by
the algebraic structure ${\cal THREE}$, shown in the double-Hasse
diagram of Figure \ref{fig:three}. Intuitively, the elements $t$
and $f$ in ${\cal THREE}$ correspond to the usual classical
elements {\tt true} and {\tt false}, while the third element,
$\top$, represents inconsistent information (or belief).

Viewed horizontally, ${\cal THREE}$ is a complete lattice. We
denote the meet, join, and the order reversing operation on the
corresponding order relation (i.e., $\leq_{t}$) by $\wedge$,
$\vee$, and $\neg$ (respectively). Viewed vertically, ${\cal
THREE}$ is a semi-upper lattice. We denote by $\oplus$ the meet
operation w.r.t. the corresponding order ($\leq_{k}$). We note
that ${\cal THREE}$ is the algebraic structure that defines the
semantics of several three-valued formalisms, such as LFI
\cite{CM01} and LP \cite{Pr89,Pr91}.

The various semantic notions are defined on ${\cal THREE}$ as
natural generalizations of similar classical ones: a {\em
valuation\/} $\nu$ is a function that assigns a truth value in
${\cal THREE}$ to each atomic formula. Any valuation is extended
to complex formulae in the obvious way. The set of the {\em
designated\/} truth values in ${\cal THREE}$ (i.e., those
elements in ${\cal THREE}$ that represent true assertions)
consists of $t$ and $\top$. A valuation $\nu$ {\em satisfies\/} a
formula $\psi$ iff $\nu(\psi)$ is designated. A valuation that
assigns a designated value to every formula in a theory ${\cal
T}$ is a (three-valued) {\em model\/} of ${\cal T}$.

Next we characterize the repairs of a database $\DB$ by its three-valued
models:

\begin{proposition}
\label{prop:repair_1}
 Let $\xDB$ be a database and let $M$ be a two-valued model of ${\cal IC}$.
 Consider the three-valued valuation $N$, defined for every atom $p$ by
 $N(p) \!=\! {\cal H}^{\cal D}(p) \oplus M(p)$, and let $\xIn  =  N^{\top}
 \setminus {\cal D}$, $\xOut = N^{\top} \cap {\cal D}$. Then
    $N$ is a three-valued model of ${\cal D} \cup {\cal IC}$, and
    $(\xIn,\xOut)$ is a repair of $\xDB$.
\end{proposition}

\noindent{\em Proof:} For the first claim, note that for
three-valued valuations $\nu$ and $\mu$, if for every atom $p$,
$\nu(p) \!\geq_{k}\! \mu(p)$, then for every formula $\psi$,
$\nu(\psi) \!\geq_{k}\! \mu(\psi)$ (the proof is by an easy
induction on the structure of $\psi$). We denote this fact by
$\nu \!\geq_{k}\! \mu$. Note also, that if $\nu \!\geq_{k}\! \mu$
and $\mu$ is a model of some theory ${\cal T}$, then $\nu$ is also
a model of ${\cal T}$. Now, since by the definition of $N$, $N
\!\geq_{k}\! {\cal H}^{\cal D}$, and since ${\cal H}^{\cal D}$ is
a model of ${\cal D}$, $N$ is a model ${\cal D}$. Similarly, $N
\!\geq_{k}\! M$, and $M$ is a model of ${\cal IC}$, thus $N$ is
also a model of ${\cal IC}$.

For the second part one has to show that the three conditions of
Definition \ref{def:repair} are satisfied. Indeed, the first two
conditions obviously hold. For the last condition, note that
${\cal D} \cup \xIn \setminus \xOut =
 {\cal D} \cup (N^{\top} \setminus {\cal D}) \setminus
 (N^{\top} \cap {\cal D}) =
 {\cal D} \cup (M^{t} \setminus {\cal D}) \setminus (M^{f} \cap {\cal D}) =
 {\cal D} \cup (M^{t} \setminus {\cal D}) \setminus ({\cal D} \setminus M^{t})
 = M^{t}$.
It follows that $M$ is the minimal Herbrand model of ${\cal D}
\cup \xIn \setminus \xOut$ and it is also a model of ${\cal IC}$,
therefore
${\cal D} \cup \xIn \setminus \xOut \models {\cal IC}$. \hfill$\Box$

\bigskip

Again, it is possible to show that the converse is also true:

\begin{proposition}
\label{prop:repair_2}
Let $(\xIn,\xOut)$ be a repair of a database $\xDB$. Then there is a
three-valued model $N$ of ${\cal D} \cup {\cal IC}$, such that $\xIn  =
N^{\top} \setminus {\cal D}$ and $\xOut = N^{\top} \cap {\cal D}$.
\end{proposition}

\noindent{\em Outline of proof:} Consider a valuation $N$, defined
as follows: \vspace{2mm} \\
\hspace*{20mm} $N(p) = \left\{ \begin{array}{ll}
     \top  \ \ & \mbox{ if $p \!\in\! \xIn \cup \xOut$, }  \vspace{1mm} \\
     t     \ \ & \mbox{ if $p \!\not\in\! \xIn \cup \xOut$ but
                        $p \!\in\! {\cal D}$, }  \vspace{1mm} \\
     f     \ \ & \mbox{ otherwise. } \\
\end{array} \right.$ \vspace{2mm} \\
Clearly, $N$ is a (three-valued) model of ${\cal D}$ and ${\cal
IC}$, and $N^{\top} \setminus {\cal D} = (\xIn \cup \xOut)
\setminus {\cal D} = \xIn$, $N^{\top} \cap {\cal D} =
(\xIn \cup \xOut) \cap {\cal D} = \xOut$. \hfill$\Box$

\bigskip

The last two propositions characterize the repairs of $\UDB$ in
terms of pairs that are associated with three-valued models of
${\cal D} \cup {\cal IC}$. We shall denote the elements of these
pairs as follows:

\begin{definition}
{\rm Let $N$ be a three-valued model and let $\DB = \xDB$ be a
knowledge-base. Denote: $\xIn^{N} = N^{\top} \setminus {\cal D}$
and $\xOut^{N} = N^{\top} \cap {\cal D}$.
} \end{definition}

We conclude this model-theoretic analysis by characterizing the
set of the $\leq$-preferred repairs, where $\leq$ is one of the
preference criteria, considered in Definition
\ref{preference_criteria} (i.e., set inclusion or differences in
cardinality).

\begin{definition}
{\rm Given a knowledge-base $\DB = \xDB$, denote: \vspace{1mm} \\
 \hspace*{5mm}
 ${\cal M}^{\DB} = \{ N ~|~ N \geq_{k} {\cal H}^{\cal D} \oplus M, \
                      M \mbox{ is a classical model of } {\cal IC} \}.$
 \footnote {Note that $N$ is a {\em three-valued\/} valuation
 and $M$ is a {\em two-valued\/} model of ${\cal IC}$. }
} \end{definition}

\begin{example}
\label{example:M^DB}
{\rm
 In what follows we shall write $M =\{p_{i} \!:\! x_{i}\}$
 for $M(p_{i}) = x_{i}$ ($x_{i} \!\in\! \{t,f,\top\}$, $i \!=\!
 1,\ldots,n$). Let $\DB = (\{p,r\},\;\{p \rightarrow q\})$. We have that
 ${\cal H}^{{\cal D}} = \{p\!:\!t, \: q\!:\!f, \: r\!:\!t\}$, and so
 ${\cal M}^{\DB} = \{ N ~|~ N(p) \!\geq_{k}\! t, \:
                              N(q) \!=\! \top, \: N(r) \!\geq_{k}\! t \}
                   \ \cup \
                   \{ N ~|~ N(p) \!=\! \top, N(q) \!\geq_{k}\! f, \:
                              N(r) \!\geq_{k}\! t \}.$
} \end{example}

\begin{definition}
{\rm Let ${\cal S}$ be a set of three-valued valuations, and
$N_{1}, N_{2} \!\in\! {\cal S}$.
\begin{itemize}
   \item $N_{1}$ is {\em $\leq_{i}$-more consistent\/} than $N_{2}$, if
         $N_{1}^{\top} \subset N_{2}^{\top}$.
   \item $N_{1}$ is {\em $\leq_{c}$-more consistent\/} than $N_{2}$, if
         $\#(N_{1}^{\top}) < \#(N_{2}^{\top})$. \footnote
         {Recall that $\#(S)$ denotes the size of $S$.}
   \item $N \!\in\! {\cal S}$ is
         {\em $\leq_{i}$-maximally consistent\/} in ${\cal S}$
         (respectively, $N$ is {\em $\leq_{c}$-maximally consistent\/}
         in ${\cal S}$), if there is no $N' \!\in\! {\cal S}$ that is
         $\leq_{i}$-more consistent than $N$ (respectively, no
         $N'\!\in\! {\cal S}$ is $\leq_{c}$-more consistent than $N$).
\end{itemize}
} \end{definition}

\begin{proposition}
\label{prop:repair_3}
If $N$ is a $\leq_{i}$-maximally consistent
element in ${\cal M}^{\DB}$, then \\
$(\xIn^{N},\;\xOut^{N})$ is a
$\leq_{i}$-preferred repair of $\DB$.
\end{proposition}

\begin{proposition}
\label{prop:repair_4}
 Suppose that $(\xIn,\;\xOut)$ is a $\leq_{i}$-preferred repair
 of $\DB$. Then there is a $\leq_{i}$-maximally consistent element
 $N$ in ${\cal M}^{\DB}$ s.t. $\xIn = \xIn^{N}$ and $\xOut = \xOut^{N}$.
\end{proposition}

\begin{note}
{\rm Propositions \ref{prop:repair_3} and \ref{prop:repair_4} hold
also when $\leq_{i}$ is replaced by $\leq_{c}$.
} \end{note}

\begin{example}
{\rm Consider again Example \ref{example:set_inclusion}. We have
that: \vspace{1mm} \\ \hspace*{5mm}
 $\UDB = \xDB = ( \, \{ p(a), \:p(b), \: q(a), \: q(c) \}, \,
                \{ \forall X (p(X) \!\rightarrow\! q(X)) \} \, ).$
\vspace{1mm} \\
Thus ${\cal H}^{{\cal D}} = \{ p(a)\!:\!t, \:p(b)\!:\!t, \:
p(c)\!:\!f, \: q(a)\!:\!t, \: q(b)\!:\!f, \: q(c)\!:\!t \}$, and
the classical models of ${\cal IC}$ are those in which either
$p(y)$ is false or $q(y)$ is true for every $y \!\in\!
\{a,b,c\}$. Now, since in ${\cal H}^{{\cal D}}$ neither $p(b)$ is
false nor $q(b)$ is true, it follows that {\em every\/} element
in ${\cal M}^{\UDB}$ must assign $\top$ either to $p(b)$ or to
$q(b)$. Hence, the $\leq_{i}$-maximally consistent elements in
${\cal M}^{\UDB}$ (which in this case are also the
$\leq_{c}$-maximally consistent elements in ${\cal M}^{\UDB}$)
are the following:
\vspace{1mm} \\
\hspace*{11mm} $M_{1} = \{ \: p(a)\!:\!t, \: p(b)\!:\!\top,
                          \: p(c)\!:\!f, \: q(a)\!:\!t, \
                          \: q(b)\!:\!f, \: q(c)\!:\!t \: \}$
\vspace{1mm} \\
\hspace*{11mm} $M_{2} = \{ \: p(a)\!:\!t, \: p(b)\!:\!t,
                          \: p(c)\!:\!f, \: q(a)\!:\!t,
                          \: q(b)\!:\!\top, \: q(c)\!:\!t \: \}$
\vspace{1mm} \\
 By Propositions \ref{prop:repair_3} and
\ref{prop:repair_4}, then, the $\leq_{i}$-preferred repairs of
$\UDB$ (which are also its $\leq_{c}$-preferred repairs) are
$(\xIn^{M_{1}},\;\xOut^{M_{1}}) = (\emptyset,\:\{p(b)\})$ and
\\
$(\xIn^{M_{2}},\;\xOut^{M_{2}}) = (\{q(b)\},\: \emptyset)$ (cf.
Example \ref{example:set_inclusion}).

Similarly, the $\leq_{i}$-maximally consistent (and the
$\leq_{c}$-maximally consistent) elements in ${\cal M}^{\DB}$,
where $\DB$ is the database of Example \ref{example:M^DB}, are
$N_{1} = \{ \: p\!:\! t, \: q\!:\! \top, \: r \!:\!t \: \}$ and
$N_{2} = \{ \: p\!:\! \top, q\!:\! f, \: r \!:\! t \: \}$. It
follows that the preferred repairs in this case are $(\{q\},\:
\emptyset)$ and $(\emptyset,\:\{p\})$.
} \end{example}

\section {Database repair -- An abductive approach}

In \cite{AvNDB01} we have presented an abductive approach to the
problem of combining inconsistent databases. In this section we
give an outline of this method. For more detailed description the
reader is referred to \cite{AvNDB01}; the application itself is
available at {\tt http://www.cs.kuleuven.ac.be/$\!\sim$dtai/kt}.

A high level description of the integration problem under
consideration is given in ID-logic \cite{De00}, which is a
framework for declarative knowledge representation that extends
classical logic with inductive definitions. This logic
incorporates two types of knowledge: definitional and assertional.
Assertional knowledge is a set of first-order statements,
representing a general truth about the domain of discourse.
Definitional knowledge is a set of rules of the form $p
\!\leftarrow\! {\cal B}$, in which the head $p$ is a predicate
and the body ${\cal B}$ is a first order formula. A predicate that
appears in a head of a rule is called {\em defined\/}; a
predicate that does not occur in any head is called {\em open\/},
or {\em abducible\/}.

A {\em theory\/} ${\cal T}$ in ID-logic is therefore a pair ({\sl
Def},$\:${\sl Fol}), where {\sl Def} (the definitional knowledge)
is a set of rules as described above, and {\sl Fol} (the
assertional knowledge) is a set of first order statements. The
meaning of ${\cal T}$ is defined by the {\em extended well-founded
semantics\/} \cite{PAA91} as follows: let $M$ be an arbitrary
two-valued interpretation for the open predicates in {\sl Def}.
Once $M$ is determined, {\sl Def} becomes a standard logic
program, with a unique well-founded model \cite{vGRS91}. This
model is then a model of the whole theory ${\cal T}$ if it is
also a model of {\sl Fol}.

ID-logic is a generalization of the notion of abductive logic
programs (ALP) \cite{DK00}. For instance, the open predicates of
a theory in ID-logic correspond to the abducibles in an abductive
logic program. Consequently, solutions of abductive logic
programs that are computed by an abductive solver are also models
of the corresponding ID-logic theory. Here we use such a solver,
called the \asystem \cite{AvNDB01,KvND01} for computing
solutions. The main idea of this solver is to reduce a high level
specification into a lower level constraint store, which is
managed by a constraint solver. The solver combines the
refutation procedures SLDNFA \cite{DD98} and ACLP \cite{KMM00},
and uses an improved control strategy. In our case, solutions are
repairs of a database, and in order to compute {\em preferred\/}
solutions (i.e., preferred repairs for the integrated database),
the \asystem has been extended with a simple branch and bound
component, called {\em optimizer\/} (see \cite{AvNDB01}). This is
actually a ``filter'' on the solutions space that speeds-up
execution and makes sure that only the desired solutions will be
obtained.

The elements of the distributed databases are uniformly
represented by the unary predicate {\tt db}, and the elements of
a repaired database are represented by the unary predicate {\tt
fact}. In order to compute these elements, two open predicates
are used: {\tt retract} and {\tt insert}. These predicates
represent, respectively, the facts that may be removed and those
that may be introduced for restoring the consistency of the
unified database. The rules for computing the elements of a
repaired database are then defined as follows: \vspace{1mm}

\noindent
\hspace*{5mm} {\tt fact(X) :- db(X), not retract(X). } \\
\hspace*{5mm} {\tt fact(X) :- insert(X). } \vspace{2mm}

In addition, the following integrity constraints are specified:
\footnote {In what follows we use the notation ``{\tt ic :- {\cal
B}}'' to denote the denial ``{\tt false $\leftarrow$ {\cal B}}''.}
\begin{itemize}
  \item It is inconsistent to have a retracted element that does not
        belong to some database: \\
        {\tt ic :- retract(X), not db(X). }
  \item It is inconsistent to have an inserted element that belongs
        to a database: \\
        {\tt ic :- insert(X), db(X). }
\end{itemize}

To make sure that all the integrity constraints will hold w.r.t.
the combined data, every occurrence of a database fact {\tt R(x)}
in some integrity constraint is replaced by ${\tt fact}(R(x))$.

Below is a code for implementing Example \ref{example:teachers}:
\footnote{The code for Example \ref{example:set_inclusion} is
similar.}

{\small
\begin{verbatim}
   defined(fact(_)). defined(db(_)). open(insert(_)). open(retract(_)).

   fact(X) :- db(X), not(retract(X)).
   fact(X) :- insert(X).
   ic :- insert(X), db(X).
   ic :- retract(X), not db(X).

   db(teaches(1,1)).  db(teaches(2,2)).                    % D1
   db(teaches(2,3)).                                       % D2
   ic :- fact(teaches(X,Y)), fact(teaches(X,Z)), Y\=Z.     % IC
\end{verbatim}
}

We have executed this code as well as other examples from the literature in
our system. The soundness and completeness theorems given in the next section
guarantee that the output in each case is indeed the set of the
most preferred solutions of the corresponding problem.

\section{Soundness and Completeness}

In this section we relate the two approaches
of the previous sections through soundness and completeness
theorems. For that we first recall some related results from
\cite{AvNDB01} (Propositions \ref{LPAR_soundness_1} --
\ref{LPAR_completeness_2} below). In what follows we denote by
${\cal T}$ an abductive theory, constructed as described in
Section 4 for defining a composition problem of $n$ databases
$\DB_{1}, \ldots, \DB_{n}$.

\begin{proposition}
\label{LPAR_soundness_1}
 Every abductive solution that is obtained by the \asystem\
 for ${\cal T}$ is a repair of $\UDB$.
\end{proposition}

\begin{proposition}
\label{LPAR_completeness_1}
 Suppose that the query `$\leftarrow {\tt true}$' has a finite
 SLDNFA-tree w.r.t. ${\cal T}$. Then every repair of $\UDB$ is obtained
 by running ${\cal T}$ in the \asystem.
\end{proposition}

\begin{proposition}
\label{LPAR_soundness_2}
 Every output that is obtained by running ${\cal T}$ in the \asystem\
 together with an $\leq_{i}$-optimizer {\rm[}respectively, together with
 a $\leq_{c}$-optimizer{\rm]} is an $\leq_{i}$-preferred repair
 {\rm[}respectively, a $\leq_{c}$-preferred repair{\rm]} of $\UDB$.
\end{proposition}

\begin{proposition}
\label{LPAR_completeness_2}
 Suppose that the query `$\leftarrow {\tt true}$' has a finite
 SLDNFA-tree w.r.t. ${\cal T}$. Then every $\leq_{i}$-preferred
 repair {\rm[}respectively, every $\leq_{c}$-preferred repair{\rm]}
 of $\UDB$ is obtained by running ${\cal T}$ in the \asystem
 together with an $\leq_{i}$-optimizer {\rm[}respectively, together
 with a $\leq_{c}$-optimizer{\rm]}.
\end{proposition}

By the propositions above and those of Section 3, we have:

\begin{corollary}
\label{sound_complete_1}
 Suppose that the query `$\leftarrow {\tt true}$' has a finite SLDNFA
 refutation tree w.r.t. ${\cal T}$. Then:
 \begin{enumerate}
    \item for every output $(\xIn,\;\xOut)$ of the \asystem for
          ${\cal T}$, there is a classical model $M$ of ${\cal IC}$
          s.t. $\xIn = M^{t} \setminus {\cal D}$ and $\xOut = {\cal D}
          \setminus M^{t}$.
    \item for every two-valued model $M$ of ${\cal IC}$ there is
          an output $(\xIn,\;\xOut)$ of the \asystem for ${\cal T}$,
          s.t. $\xIn = M^{t} \setminus {\cal D}$ and $\xOut =
          {\cal D} \setminus M^{t}$.
 \end{enumerate}
\end{corollary}

\begin{corollary}
\label{sound_complete_2}
 Under the same assumption as that of Corollary \ref{sound_complete_1},
 \begin{enumerate}
    \item for every output $(\xIn,\;\xOut)$ of the \asystem for ${\cal T}$
          there is a 3-valued model $N$ of ${\cal D} \cup
          {\cal IC}$, s.t. $\xIn^{N} \!=\! \xIn$ and $\xOut^{N} \!=\! \xOut$.
    \item for every 3-valued model $N$ of ${\cal D} \cup {\cal IC}$ there is
          an output $(\xIn,\;\xOut)$ of the \asystem for ${\cal T}$, s.t.
          $\xIn \!=\! \xIn^{N}$ and $\xOut \!=\! \xOut^{N}$.
 \end{enumerate}
\end{corollary}

\begin{corollary}
\label{sound_complete_3}
 In the notations of Corollary \ref{sound_complete_1} and under its
 assumption,
 \begin{enumerate}
    \item for every output $(\xIn,\;\xOut)$ that is obtained by
          running ${\cal T}$ as an input to the \asystem together
          with an $\leq_{i}$-optimizer {\rm[}respectively, together with a
          $\leq_{c}$-optimizer{\rm]}, there is an $\leq_{i}$-maximally
          consistent element {\rm[}respectively, a $\leq_{c}$-maximally
          consistent element{\rm]} $N$ in ${\cal M}^{\UDB}$ s.t.
          $\xIn^{N} = \xIn$ and $\xOut^{N} = \xOut$.
    \item for every $\leq_{i}$-maximally consistent element
          {\rm[}respectively, $\leq_{c}$-maximally consistent element{\rm]}
          $N$ in ${\cal M}^{\UDB}$ there is a solution $(\xIn,\;\xOut)$ that
          is obtained by running ${\cal T}$ in the \asystem together
          with an $\leq_{i}$-optimizer {\rm[}respectively, together with
          a $\leq_{c}$-optimizer{\rm]} s.t. $\xIn = \xIn^{N}$
          and $\xOut = \xOut^{N}$.
 \end{enumerate}
\end{corollary}

\section{Related works}

Coherent integration and proper representation of amalgamated
data is extensively studied in the literature (see, e.g.,
\cite{BKM91,Br97,FPLPS01,GL00,GZ00,LS00,LM98,Me97,Ol91,Re93,Su94}).
Common approaches for dealing with this task are based on
techniques of belief revision \cite{LS00}, methods of resolving
contradictions by quantitative considerations (such as ``majority
vote'' \cite{LM98}) or qualitative ones (e.g., defining
priorities on different sources of information or preferring
certain data over another \cite{Ar99,BCDLP93}), and approaches
that are based on rewriting rules for representing the
information in a specific form \cite{GZ00}. As in our case,
abduction is used for database updating in \cite{KM90} and an
extended form of abduction is used in \cite{IS95,SI99} to explain
modifications in a theory.

The use of three-valued logics is also a well-known technique for
maintaining incomplete or inconsistent information; such logics
are often used for defining fixpoint semantics of incomplete logic
programs \cite{Fi85,vGRS91}, and so in principle they can be
applied on integrity constraints in an (extended) clause form
\cite{De00}. Three-valued formalisms such as LFI \cite{CM01} are
also the basis of paraconsistent methods to construct database
repairs \cite{dACM02} and are useful in general for pinpointing
inconsistencies \cite{Pr91}. As noted above, this is also the
role of the three-valued semantics in our case.

Other approaches are based on semantics with arbitrarily many
truth values, which allow to decode within the language itself
some ``meta-information'' such as confidence factors, amount of
belief for or against a specific assertion, etc. These approaches
combine corresponding formalisms of knowledge representation
(such as annotated logic programs \cite{Su90,Su94} or
bilattice-based logics \cite{AA96,Fi91,Me97}) together with
non-classical refutation procedures \cite{Fi89,KL92,Su90} that
allow to detect inconsistent parts of a database and maintain
them.

A closely related topic is the problem of giving consistent query
answers in inconsistent database \cite{ABC99,BDP95,GZ00}. The
idea is to answer database queries in a consistent way {\em
without\/} computing the repairs of the database.

There are some other applications for integrating possibly
conflicting information and updating databases (e.g., LUPS
\cite{ALPQ00}, BReLS \cite{LS00}, RI \cite{KL92}, Subrahmanian's
mediator of annotated databases \cite{Su94}, and the system of
Franconi et al. \cite{FPLPS01}). In comparison with such systems, we note
that the main advantages of the present application are its
expressive power (to the best of our knowledge, our approach is
more expressive than any other available application for coherent
data integration), the fact that no syntactical embedding of
first-order formulae into other languages nor any extensions of
two-valued semantics are necessary (our approach is a pure
generalization of classical refutation procedures), and the
encapsulation of the way that the underlying data is kept coherent
(no input from the reasoner nor any other external policy for
making preferences among conflicting sources is compulsory in
order to resolve contradictions).

\section{Future work}

We conclude by sketching some issues for future work.
First, as we have already noted, two more phases, which have not
been considered here, might be needed for a complete data
integration: (a) translation of difference concepts to a unified
ontology, and (b) resolving contradictions among different
integrity constraints. Another issue for future work is to allow
definitions of {\em concepts\/} (and not only integrity
constraints) in the databases (see \cite{De00} for a sketch on how
this may be done). This data may be further combined with
(possibly inconsistent) temporal information, (partial)
transactions, and (contradictory) update information. Finally,
since different databases may have different information about
the same predicate, it is reasonable to use some weakened version
of the closed word assumption as part of the integration process
(for instance, an assumption that something is false unless it is
in the database, or some other database has some information
about it). An alternative approach may be to replace the closed
word assumption with partial valuations (in case that databases
may contain negative facts and not only positive ones).

\end{document}